\documentclass[preprint,tighten,floats,twoside,eqsecnum,aps,epsf,epsfig]{revtex4}

\usepackage{amssymb}

\usepackage{graphicx}

\usepackage{color}
\usepackage{bm}
\newcommand{\GeV}{\makebox{ GeV}}
\newcommand{\beq}{\begin{equation}}
\newcommand{\enq}{\end{equation}}
\newcommand{\beqa}{\begin{eqnarray}}
\newcommand{\beqast}{\begin{eqnarray*}}
\newcommand{\enqa}{\end{eqnarray}}
\newcommand{\enqast}{\end{eqnarray*}}
\newcommand{\nn}{\nonumber}

\newcommand{\ga}{\gamma}
\newcommand{\de}{\delta}
\newcommand{\ep}{\epsilon}

\newcommand{\si}{\sigma}

\newcommand{\om}{\omega}

\newcommand{\Ga}{\Gamma}

\def\GeV{\nobreak\,\mbox{GeV}}

\begin{document}
\title{Universality in the Electroproduction of Vector Mesons}
\author{V.L. Baltar}
\affiliation{Departamento de F\'{\i}sica, Pontif\'{\i}cia Universidade Cat\'olica do Rio de 
 Janeiro \\  
Rio de Janeiro 22453-900, RJ, Brazil }
\author{H.G. Dosch } 
 \affiliation{Institut f\"ur Theoretische Physik, 
  Philosophenweg 16, D-6900 Heidelberg, Germany }
 \author{E. Ferreira}
 \affiliation{Instituto de F\'{\i}sica, Universidade Federal do Rio de
 Janeiro \\
 C.P. 68528, Rio de Janeiro 21941-972, RJ, Brazil   }
 
\begin{abstract}

We study universality in the electroproduction of vector mesons using a unified 
nonperturbative approach which has already proved to reproduce extremely well the 
available experimental data. In this framework, after the extraction of factors 
that are specific of each vector meson, we arrive at a reduced integrated elastic 
cross section which is universal. Our calculations suggest a finite infrared behavior 
for the strong coupling constant.
\end{abstract}
 
\bigskip
  \maketitle

\section{Introduction}\label{intro}
 
Experiments show a striking similarity among the production cross sections of 
the various vector mesons, pointing to the existence of a universal content 
in these processes.  In the present work we study this universality property 
using a nonperturbative procedure which describes in a unified way the 
cross sections for elastic electroproduction of the S-wave vector mesons.

We employ a general treatment of high energy scattering based on the functional 
integral approach to QCD~\cite{Nachtmann,Nachtmann05,Nachtmann07} and on the 
WKB method. The functional integrals are calculated nonperturbatively, in an 
extended stochastic vacuum model~\cite{Dosch,DoschSimonov}. This nonperturbative
approach has been successfully applied in many fields, from hadron spectroscopy 
to high energy scattering \cite{DFK94,DDLN02,DDSS02}, DVCS \cite{DD01} and  
structure functions \cite{DD02}. Recently, the method has been shown to provide a 
good description of the existing observed data on vector meson electroproduction 
and a comparison with the characteristic features of the perturbative approaches 
~\cite{Donnachie,ryskin93,brodsky94,ryskin97,strikman97,strikman98,maor,nikolaev}
has been extensively discussed \cite{DF07}.

In our scheme the structure of the transition amplitude is defined in terms of 
color neutral quark-antiquark states (color dipoles) that displace themselves 
in the external QCD vacuum field. The fundamental elementary dynamical 
effect is given by the loop-loop interaction: Wilson loops formed by dipoles in the 
photon-meson overlap product and in the proton (for simplicity treated 
in a diquark model). The transition to observable electroproduction amplitudes 
of hadrons is obtained through a superposition of the loop-loop amplitudes
with the light cone wave functions of the hadrons and the photon used as weights.
This procedure allows the calculation of polarized transition amplitudes and angular 
distributions within a unified view of all processes of vector meson electroproduction.  
We need no external inputs, and the only intervening quantities are inherently 
calculated in the nonperturbative approach; no new free parameters have to be 
introduced. A more detailed description of the method can be found in \cite{DFK94}.

In order to look for the structure of the universal behavior that appears in the data 
we concentrate on the photon virtuality $ ( Q^2 ) $  dependence of the integrated 
elastic cross sections. This is the simplest observed quantity, determined almost 
fully by the shape of the photon and vector meson wave functions, both described 
as packets of quark-antiquark dipoles. Universality is inherent to the nonperturbative 
nature of the method here applied, and is reproduced accurately by our theoretical 
calculations.

We first consider the loop-loop amplitude which depends on the dipole sizes 
and orientation and on the impact parameter. The amplitude for the hadronic 
process in the impact parameter space is obtained by integrating the square of 
the loop-loop amplitude over the orientation and sizes of the dipoles, 
with the above mentioned weight factors, i.e., the squares of the proton wave 
function and the square of the overlaps of photon and meson wave functions. 
A Fourier transform leads from the impact parameter picture in the transverse plane 
to the momentum transfer description. After integration over the geometric weighted 
distribution of loops, the correct $Q^2$ dependence and normalization 
of cross sections emerge \cite{EFVL} , which turns out to have form 
      $ \sigma={A_V}/{(1+Q^2/M_V^2)^n} ~ , $   where $M_V$  is the vector meson mass.
We notice that this essential variable arises naturaly in our calculations, 
in spite of $M_V$ not being introduced a priori in this framework. Actually the 
vector meson mass appears  numerically only through the expression, given in the next 
section, which relates the electromagnetic coupling $f_V$ to the measured decay 
rate $\Ga_{V\rightarrow e^+e^-}$.

The numerical factors $A_V$  that fix the normalization of the cross sections, 
as well as the power $n$,  are determined uniquely by the model, without free 
parameters, reproducing the data with good precision. The values of $n$ differ little 
from one vector meson to another\cite{EFVL}. Predicted cross sections 
cover a range of 5 orders of magnitude \cite{DF07}. 

The previously known property regarding universality is that, divided by the square 
of the quark charges of the vector mesons, (in the $\rho$  and  $\omega$  
meson cases, formed with differents quarks u and d,  this means the average 
effective charges), the values of the constans $A_V$ are quite similar. 

We here show an important improvement of this universality 
property: instead of extracting the squared charges, we extract the 
squared electromagnetic couplings $f_V^2$. Besides leading to a closer 
grouping of the cross section values, this criterium to reach universality 
is more fundamental, as the couplings fix the properties of the wave 
functions. The electromaqnetic coupling, experimentally accessible through 
the electromagnetic decay rates, is directly related to the value of the 
wave function at the origin.  The interesting comparison of the universality 
in  cross section values obtained with extraction of squared charges and 
with extraction of squared couplings is shown in section 3. 

We then go a step further, and examine the expression that relates 
 the couplings $f_V$ to the experimental electromagnetic decay rates 
$\Ga_{V\rightarrow e^+e^-}$.
 A first order radiative correction to the decay rate  
\cite{Barbieri,Celmaster} introduces a  factor 
 $$ \left[ 1 - \left(\frac{16}{3\pi}\right) \alpha_S(m_f)\right] $$
with respect to the bare decay rate.  The situation is 
that the fundamental universality property should be found with 
extraction of the pure coupling, $f_V^{(0)}$, not of the 
effective one $f_V$, with the radiative correction factor. 

The first order radiative correction to the decay rate  depends on
the intervening quark mass, because of the running property of the 
strong coupling constant  $\alpha_S(m_f)$.  This quantity is 
known for values  of the argument $m_f$ that are larger than about 1 GeV.
Thus the bare coupling can be numerically determined for the 
charm and bottom cases. We show that, using the resulting values of bare 
couplings in $J/\psi$ and $\Upsilon$ decays, perfect matching 
of the corresponding reduced cross sections is obtained. 

For the light mesons  we have to consider effective values of the strong 
coupling constant. Studies about the infrared  regularization of the gluon propagator 
point to finite values for  $\alpha_S(0)$.  We explore the possible criterion that 
the complete universality in the cross sections fixes  effective values 
for $\alpha_S(m_f)$ , and obtain interesting results at the light quark masses.

Our paper is organized as follows. In Section (\ref{basic}) we very briefly expose 
our approach to the analysis of vector meson production. In Section (\ref{universality})
we show that, with the appropriate reductions, our calculations lead 
to a description of these processes which exhibits a universality 
determined by properties of the overlap of photon and vector meson 
wave functions. Section (\ref{conclusions}) concludes with a summary and discussion of our 
results. In the Appendix we present details of the vector meson light 
cone wave functions.

\section{Basic formul\ae~and results}	\label{basic}

For  convenience we present here  some basic formulae and 
results developed in our previous work on photo-  and 
electroproduction ~\cite{DF07,EFVL,DGKP97,KDP99,DF02,DF03}, 
where details can be found.

The amplitude for electroproduction of a  vector meson V in
polarization  state $\lambda$  is written
in our framework
\beq \label{TM}
T_{\gamma^* p \to V p,\lambda}(s,t;Q^2)= \int d^2 {\mathbf R}_1 dz_1  ~
 \rho_{\ga^* V,\lambda} (Q^2;z_1,{\mathbf R}_1)  ~
J(s,{\mathbf q},z_1,{\mathbf R}_1) ~ ,
\label{int} \enq
with
\beq
J(s,{\mathbf q},z_1,{\mathbf R}_1)= \int  d^2 {\mathbf R}_2
d^2 {\mathbf b} \, e^{-i {\mathbf q}.{\mathbf b}}
|\psi_p({\mathbf R}_2)|^2  S(s,b,z_1,{\mathbf R}_1,z_2=1/2,
  {\mathbf R}_2) ~ .
\label{int2} \enq
Here $\psi_p({\mathbf R}_2)$ is the proton wave function and
\beq \label{overlap}
 \rho_{\ga^* V,\lambda} (Q^2;z_1,{\mathbf R}_1) =
        \psi_{V\lambda}(z_1,{\mathbf R}_1)^*\psi_{\ga^* \lambda}(Q^2;z_1,{\mathbf R}_1)
\label{over}
\enq
represents the overlap of photon and vector meson wave functions;
$ S(s,b,z_1,{\mathbf R}_1,1/2,{\mathbf R}_2) ~ $   is the scattering
amplitude of two dipoles with separation vectors
${\mathbf R}_1, {\mathbf R}_2$,  colliding
with impact parameter vector  $\mathbf b$~\cite{DF02}; $\mathbf q$ is the momentum
transfer
\beq
t= (p_V - p_{\gamma}^*)^2= - {\mathbf q}^2 - m_p^2 (Q^2+M_V^2)/s^2 + O(s^{-3})\approx -{\mathbf q}^2 ~ 
\enq
In the expressions above,~$z$~is the longitudinal momentum fraction of the quark in the virtual photon and in the vector meson and $Q^2 = -{p_{\gamma}^*}^2$~is the photon virtuality.

The differential cross section is given by
\beq
\frac{d \si}{d|t|} = \frac{1}{16 \pi s^2} |T|^2 ~ .
\enq
Here,~$s$~is the center of mass energy of the~${\gamma}^*$-proton system. 

The form of overlap written in Eq.~(\ref{over}) corresponds
to SCHC (s-channel helicity conservation); generalizations can be
 made introducing a matrix in the polarization indices.
 In principle a three quark wave function should be used for the 
proton but earlier investigations have shown that a quark-diquark 
structure leads to fenomenologically consistent results and we 
adopt this model~\cite{DF02}. 

The light cone wave functions of the photon and vector meson have 
been previously discussed, and are presented in the Appendix to 
keep the paper self contained. Here we write only the scalar part 
of the meson wave function proposed by Brodsky and Lepage~\cite{lep80}, 
which is used in the present calculations: 
\beq \label{BL}
\phi_{BL}(z,r) = \frac{N}{\sqrt{4 \pi~}}
\exp\Big[-\frac{m_f^2(z-\frac{1}{2})^2}{ 2z\bar{z}\om^2 }\Big]~\exp[-2
z\bar{z}\om^2 r^2] ~ .
\enq
\smallskip
In previous papers \cite{DF02,DF03}, another  form of meson wave function
has been used : the Bauer-Stech-Wirbel (BSW) \cite{BSW85,BSW87}. Since it 
results in no essential difference for the purposes here considered, we 
concentrate on the BL wave function. 

In the expression above~$m_f$~is the quark mass and~$\bar z = 1 - z$~. 
We note that~$\phi_{BL}$~has two parameters: the normalization 
constant $N$ ( fixed by the wave function normalization) and the 
transverse size parameter~$\om$. In order to determine 
$\om$ ~\cite{DGKP97,KDP99,DF02}~we turn to the Van Royen-Weisskopf 
formula (including color)~\cite{Van} which relates the leptonic 
decay width~$\Ga_{V\rightarrow e^+e^-}$ to the wave function at 
the origin~ $\psi(0)$
\beq \label{Van}
\Ga{(V\rightarrow e^+e^-)} = \frac{16\pi\alpha^2 \hat e_V^2 |\psi(0)|^2}{M_V^2}
\enq
where $\alpha = 1 / 137.036$,  $\hat e_V$ is the effective charge 
in the vector meson, in units of the elementary charge $e$ and $M_V$ 
is the vector meson mass. This relation is the lowest order static 
approximation for the electromagnetic leptonic decay width of a 
vector meson~\cite {stan}.

This condition on the wave function, which determines  $\om$, 
is related to $f_V$, the coupling of the vector meson to the 
electromagnetic current, defined through
\beq\label{EM} 
<0|J_{em}^{\mu}(0)|V(q,{\lambda})> =  e f_V M_V {\epsilon}^{\mu}(q,\lambda)
\enq
If we use Eq.~\ref{EM} in the S matrix element 
$<e^+(p,s)e^-(p \prime, s \prime )\vert S \vert V(q,{\lambda})>$ 
and then calculate the decay rate of the vector meson, we obtain 
the relation of $f_V$ to $\Ga_{V\rightarrow e^+e^-}$~\cite{KDP99}
\beq\label{fv} 
f_V^2 = \frac{3 M_V \Ga_{V\rightarrow e^+e^-}}{4\pi\alpha^2}
\enq
>From Eqs.~\ref{fv} and \ref{Van} we then obtain
\beq\label{gamapsi} 
f_V^2 = \frac{3 M_V \Ga_{V\rightarrow e^+e^-}}{4\pi\alpha^2} = (\frac{12}{M_V})e_V^2|\psi (0)|^2
\enq
relating $f_V$ , $\Ga_{V\rightarrow e^+e^-}$ and $\psi (0)$. 

Mesons in different states of polarization are represented 
by different wave functions. Consequentely, $N$ and $\om$ 
are different for transverse and longitudinal states.

The  relations between $f_V$ and the wave functions 
are~\cite{DF02,DF07}
\beq\label{fV1} 
f_V^T = \hat e_V\frac{N}{4\pi}\frac{\sqrt 6}{M_V}\int_0^1 \frac{dz}{z\bar z}[ 2\om ^2(z^2+\bar z^2)(4z\bar z)^2~+~m_f^2]\exp\Big [-\frac{m_f^2(z-\frac{1}{2})^2}{2z\bar z\om^2}\Big ]
\enq
in the transverse case and
\beq\label{fV0} 
f_V^L = \hat e_V\frac{N}{4\pi} \sqrt {3}~ 16~ \om \int_0^1 dz~ z\bar z~ \exp\Big [-\frac{m_f^2(z-\frac{1}{2})^2}{2z\bar z\om^2}\Big ]
\enq
in the longitudinal case.

Numerically, we use $f_V^T = f_V^L = f_V$, with $f_V$ 
calculated through Eq.~(\ref{fv}). The values thus obtained, 
together with other vector meson data, are presented in 
Table~\ref {PDGtab}, in the Appendix. There, we also show 
in Table~\ref{WFparam} our results for~$N$,~$\om$,~and the 
radius for each meson in each state, calculated through 
the wave function normalization equations, Eqs.~(\ref{fV1}) ~
and~ (\ref{fV0}) ~above (with $f_V$ from Table~\ref{PDGtab}) and 
the mean value of the radius $r$. For the heavy quarks we take 
the renormalized masses $m_c$ = 1.25 GeV, $m_b$ = 4.2 GeV 
and for the light quarks we take the effective masses 
determined \cite{DGP98} as  $m_u = m_d$ =0.2, $m_s$ = 0.3 GeV.

For the quark-diquark proton wave function, fixing the 
longitudinal momentum fraction at~$z~\approx \frac{1}{2}$~ 
and making a Gaussian ansatz, we have
\beq\label{proton} 
\psi_{p}(r) = \frac{1}{2\pi} \frac{1}{S_p} \exp\Big [\frac{-r^2}{(2S_p)^2}\Big ] ~, 
\enq
where the size parameter is chosen as ~$S_p = 0.74$  fm  \cite{DNPW}.
  
In our model the energy dependence is motivated by the two-pomeron approach of Donnachie 
and Landshoff \cite{DL98}. The hard pomeron leads to $(s/s_0)^{0.42}$ for $R_1\leq r_c $ 
and  the soft pomeron leads to $({s}/{s_0})^{0.0808}$ for large dipoles, so that the 
integration over $R_1$ in Eq.~(\ref{TM}) is split into two parts \cite {DF02}. We use the 
numerical values  $r_c \approx 0.22$ fm and $s_0= (20~ GeV)^2$, taken from \cite{DonD}.
 
\section{Universality}	\label{universality}

As said above, we have used a nonperturbative QCD framework to calculate 
the observables of vector meson photo- and electroproduction, and we 
have obtained very good description of the data in all cases 
~\cite{DF07,EFVL,DGKP97,KDP99,DF02,DF03}. 
Fig. {\ref{sigmas} shows the data 
and the theoretical calculations of the electroproduction 
of $\rho, \omega, \phi, J/\psi ~{\rm and}~  \Upsilon$  vector mesons. 

The quantitative predictions made in a unique way for different 
kinds of vector mesons cover 5 orders of magnitude in the  cross 
sections~\cite{DF07}. The nonperturbative calculation leads to 
cross sections that plotted against $Q^2 + M_V^2$ are nearly 
parallel straight lines, a result confirmed by experiments. 
\begin{figure}
 \caption { Integrated elastic cross sections in vector meson 
     electroproduction: experimental data and theoretical 
 calculations  with the stochastic vector model.}  
\label{sigmas}
\includegraphics[height=10.0cm]{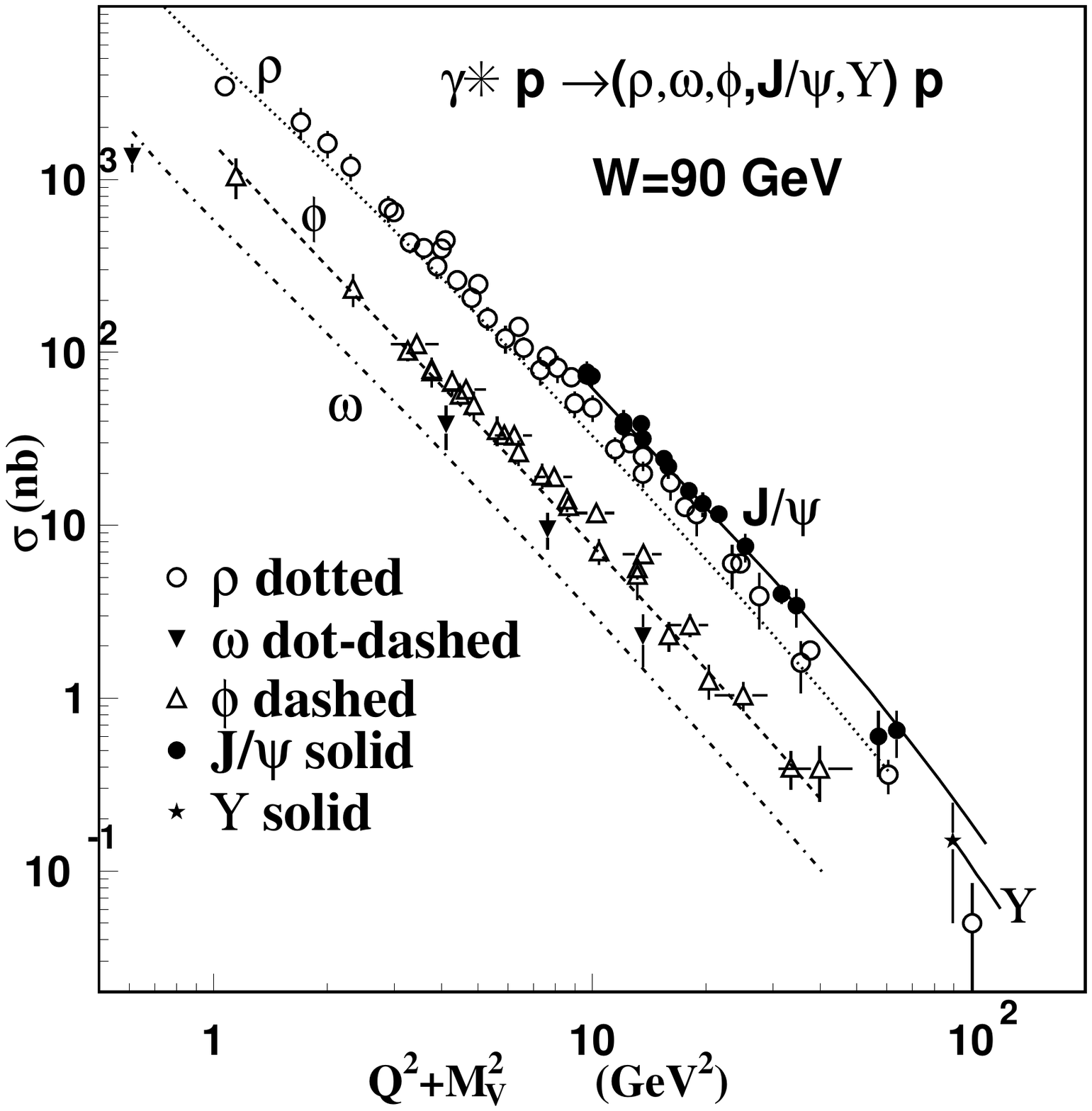}
\includegraphics[height=10.0cm]{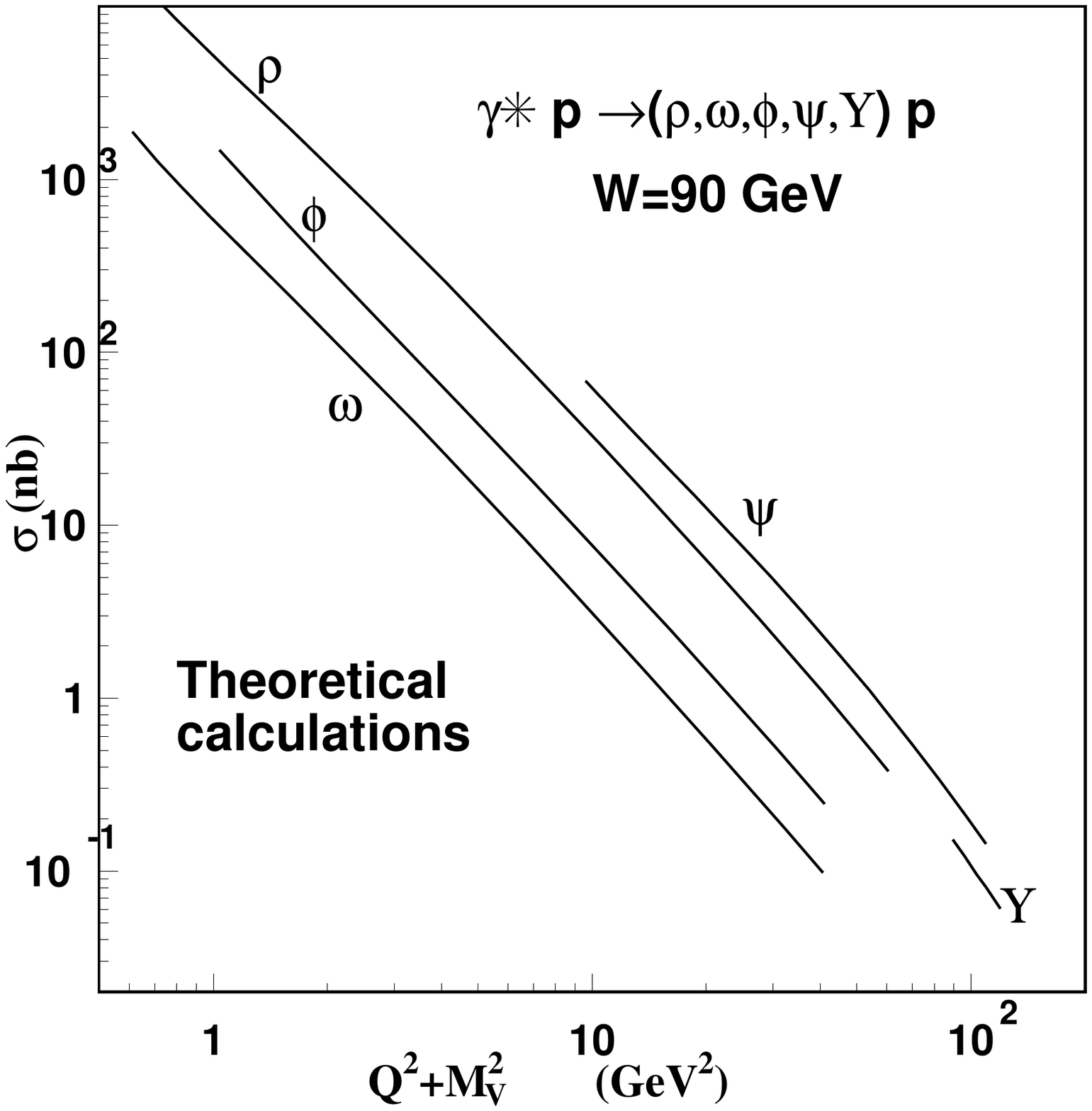}
 \end{figure}
Our work has shown that  overlap of photon and vector meson wave functions 
alone determines large part of the behavior observed in these processes. 
This simple structure  suggests a search for universality.

The effect of factorization in the magnitudes of the cross sections is 
exhibited with the extraction of charge factors 
$\hat e^2_V = 1/2 ~ , ~  1/18  ~, ~ 1/9 ~, ~ 4/9 ~, ~ 1/9 $   for 
$\rho , ~\omega ,~ \phi ,~\psi ,$ and $\Upsilon $, respectively. 
Fig. {\ref{charges_extracted} shows the theoretical results for all 
vector mesons after charge reduction, with approximately equal values 
for all cross sections. Since it can be seen from Fig. {\ref{sigmas} 
that the experimental values agree very well with the theoretical predictions
the universality to be derived from the theoretical expressions reflects 
an universality in the data.
  \begin{figure}
 \caption { Extraction of charge factors $e_V^2$  from the integrated 
   elastic cross sections in vector meson 
     electroproduction. }  
\label{charges_extracted}
\includegraphics[height=10.0cm]{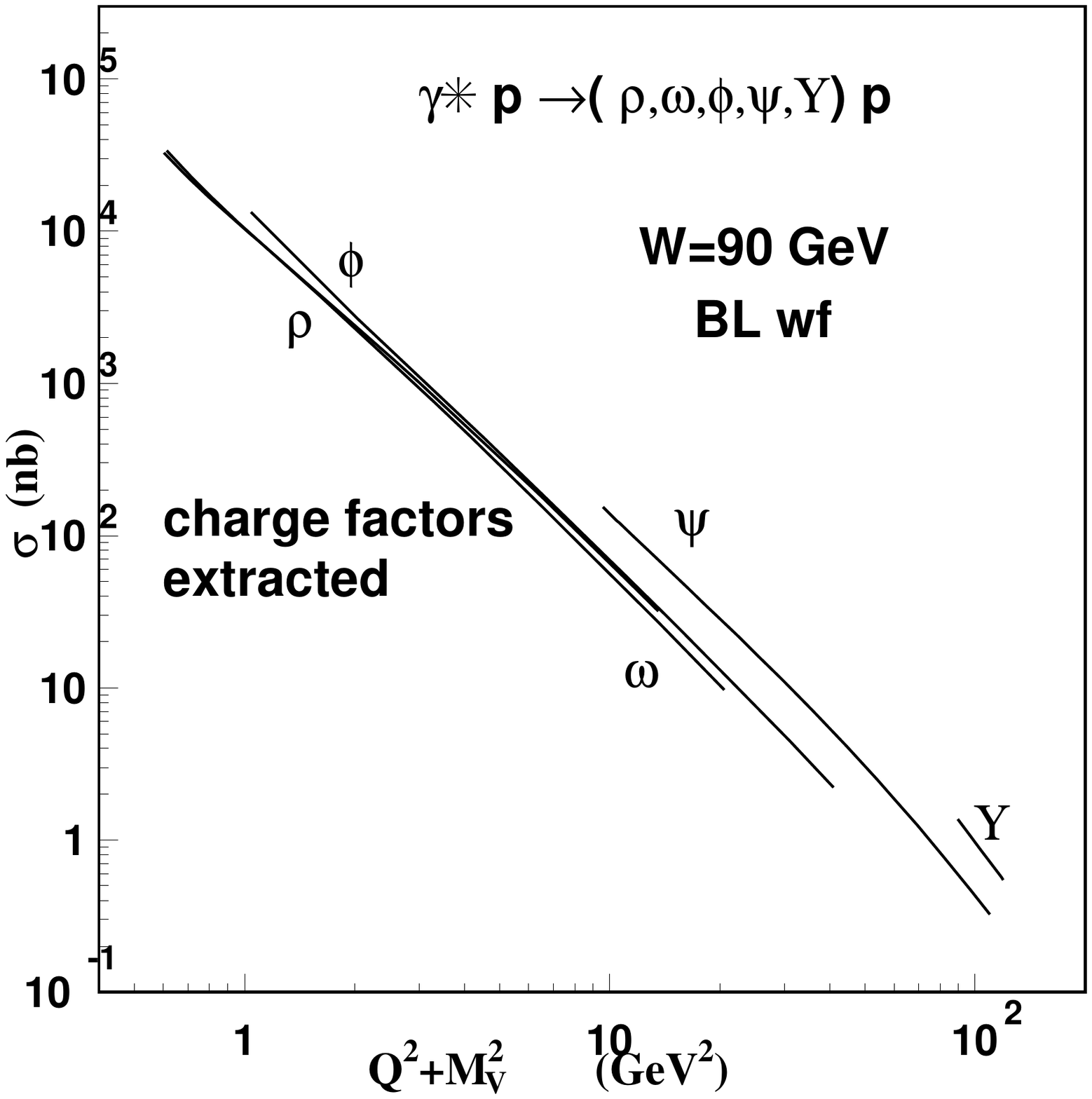}
 \end{figure} }
The agreement of our nonperturbative approach with the experimental results 
is a consequence also of the construction of the meson wave function, whose 
shapes are fixed by the  experimental values of  the  electromagnetic 
leptonic decay rates of vector mesons  $\Ga_{V\rightarrow e^+e^-}$ . 
These quantities fix the transverse size parameter $\om$ contained in 
the wave function, as explained above ~\cite{DGKP97,KDP99,DF02}.  
The coupling of the vector meson to the electromagnetic current 
($f_{V}$) is related to $\Ga_{V\rightarrow e^+e^-}$ as shown in 
Eq.~(\ref{gamapsi}). The values obtained for $f_{V}$ are collected 
in Table~\ref{PDGtab} in the Appendix.

Fig. {\ref{fV2_extracted} shows the theoretical calculations 
    of the electroproduction of $\rho$ , $\omega$, $\phi$, $J/\psi$ and 
  $\Upsilon$  vector mesons divided by the values of $f_{V}^2$.
We observe a remarkable universality, closer than obtained with 
extraction of factors of  squared charges.  The small remaining 
splittings may be due to flavor dependent corrections in the 
calculations of the electromagnetic decay rates. We take these into 
account in the discussion that follows. 
 \begin{figure}
 \caption { Extraction of $f_V^2$ coupling factors from the integrated 
   elastic cross sections in vector meson 
     electroproduction. }  
\label{fV2_extracted}
\includegraphics[height=10.0cm]{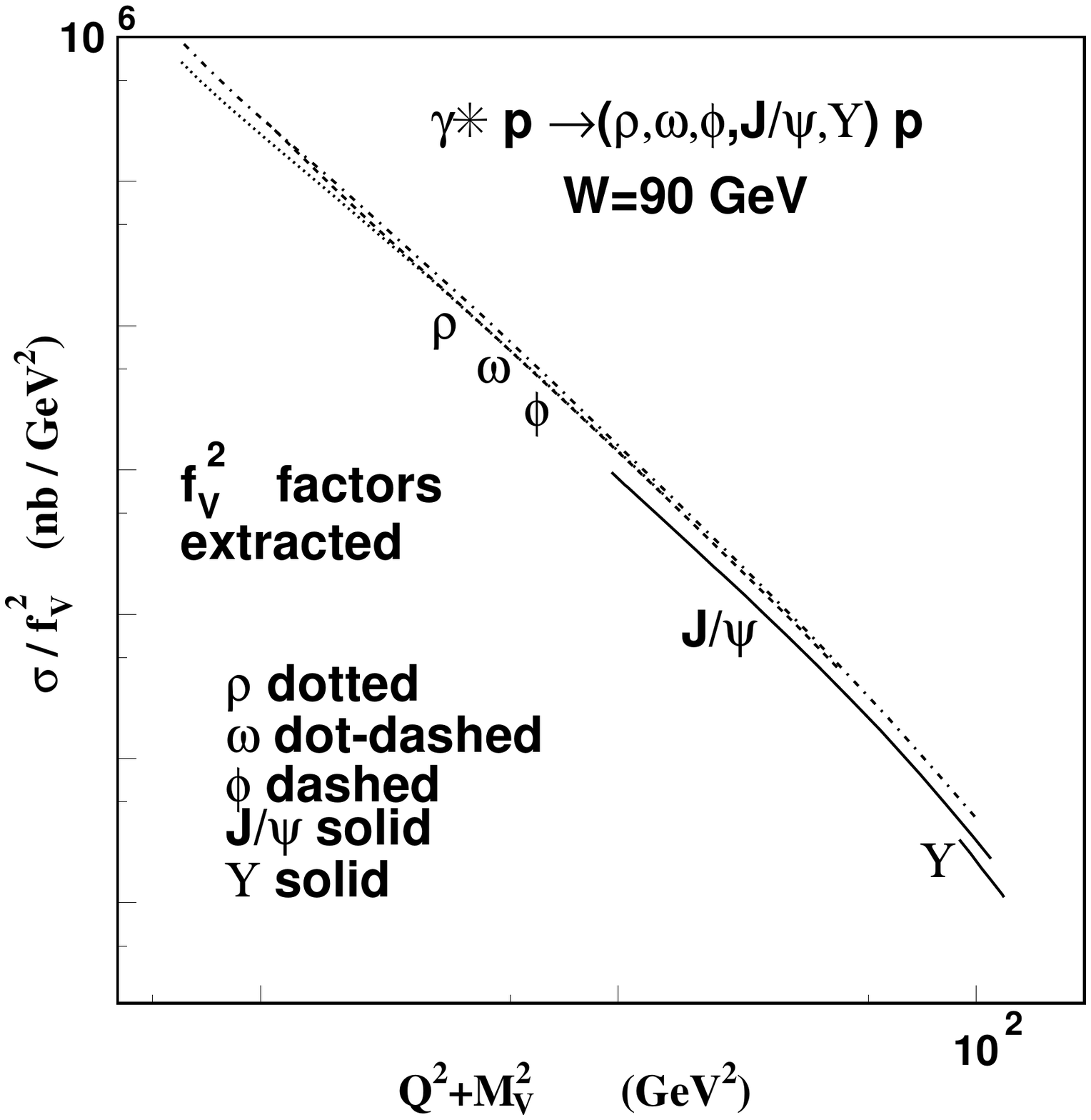}
 \end{figure}
The leading QCD radiative correction can be calculated in perturbation theory 
\cite{Barbieri,Celmaster,Tye,Bissey} and causes a significant suppression in 
the value of $\Gamma$ given by the zero-order equation (\ref{Van}), 
 which becomes
\beq \label{celmaster}
\Ga{(V\rightarrow e^+e^-)} = \Ga_0\left[ 1 - \left(\frac{16}{3\pi}\right) \alpha_S(m_f)\right]
\enq
leading to
\beq \label{corrected}
f_V^2 = 
f_V^{(0)2}F(m_f)\approx f_V^{(0)2}\left[ 1 - \left(\frac{16}{3\pi}\right) \alpha_S(m_f)\right]
\enq
The flavor dependence comes through the {\it running strong coupling constant}  
$\alpha_S$, which depends on $m_f$. 
Since $\alpha_S$ is known at the mass values of the $b$ and $c$ quark masses \cite{Hinchliffe}
\begin{center}
$\alpha_S(m_b) = 0.23\pm 0.01$~~ , ~~$\alpha_S(m_c) = 0.33\pm 0.01 ~, $
\end{center}
the corrected Eq.~(\ref{corrected}) can be applied directly to $\Upsilon$ and $J/\psi$.

Fig. {\ref{upsilon_psi} shows the theoretical calculations 
    of electroproduction of $J/\psi~ {\rm and}~
  \Upsilon$  vector mesons with extraction of $f_V^{(0)2}$ factors, 
 with remarkable coincidence of the reduced values of the cross sections. 
\begin{figure}
 \caption { Extraction of the bare $f_V^{(0)2}$ coupling factors from the 
integrated elastic cross sections of $J/\psi$  and 
  $\Upsilon$ mesons .}  
\label{upsilon_psi}
\includegraphics[height=10.0cm]{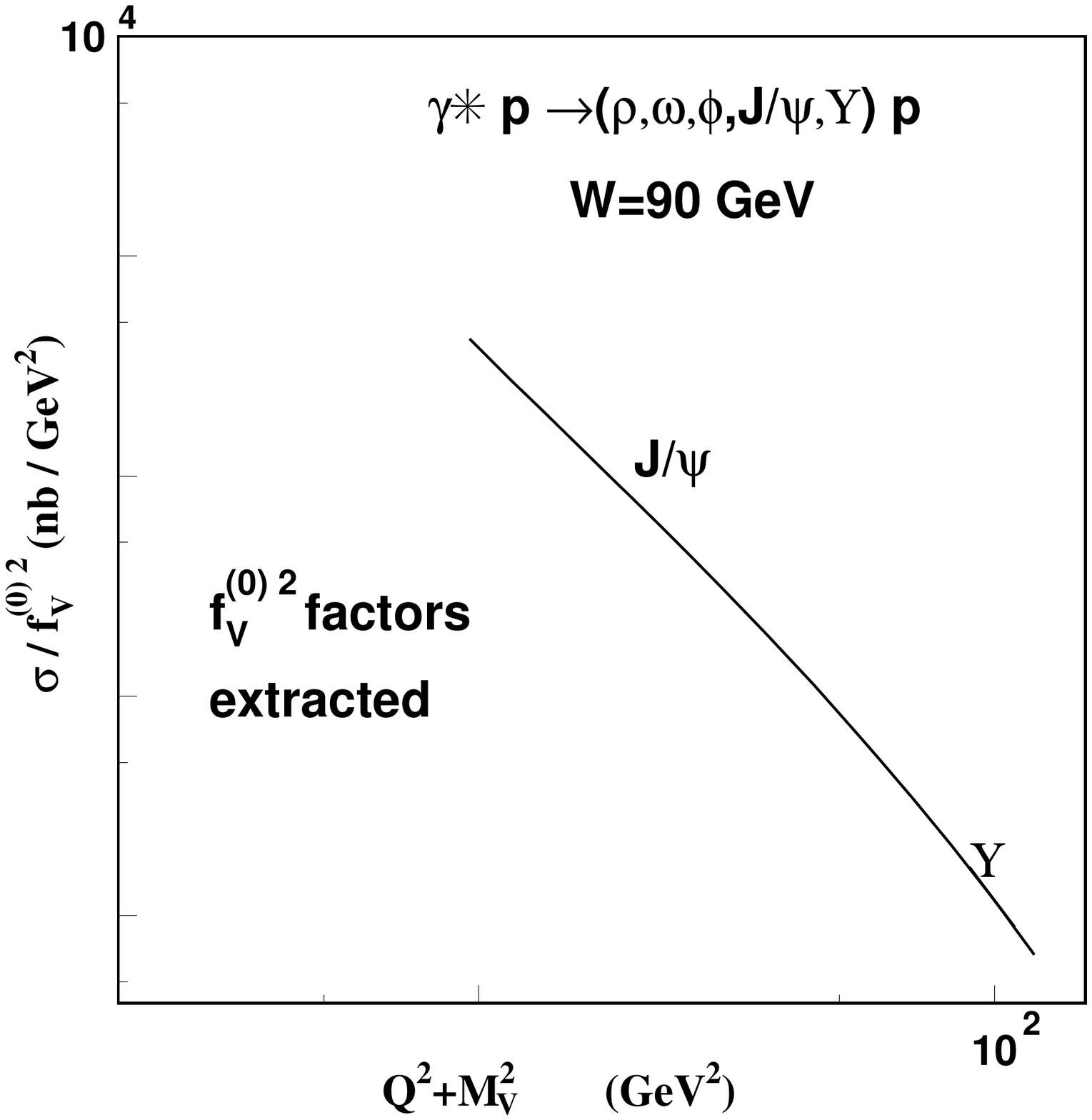}
 \end{figure}
We thus have precise universality for $\Upsilon$ and $J/\psi$ mesons.

The simple behavior connecting $J/\psi$ and $\Upsilon$ 
cross sections, that we explained by sheer acting on wave 
function construction, is a manifestation of the factorization
that shows that the whole $Q^2$ dependence of the cross
sections is determined by the overlap of the photon and 
vector meson wave functions. In Fig. \ref{pure_psi} we 
exhibit the accuracy of the calculation of the $J/\psi$
cross section using only the overlap strength of photon 
and $J/\psi$ wave functions (the overlap strength is formed by 
the integration over the internal variables of the $q \bar q$
pairs of the overlap function multiplied by $r^2$) \cite{EFVL}. 
The accurate calculation of the  cross section with
 the stochastic vacuum model in the LHS is compared to the 
calculation with  the overlap strengths in the RHS. A numerical 
factor 24000 takes into account the interaction of the proton 
with the QCD field, which has no $Q^2$ dependence.
\begin{figure}
 \caption { Data and calculation of $J/\psi$ cross section 
as a function of photon virtuality . In the RHS 
we show the $Q^2$ dependence of the overlap strength 
determined only  by wave functions.}  
\label{pure_psi}
\includegraphics[height=10.0cm]{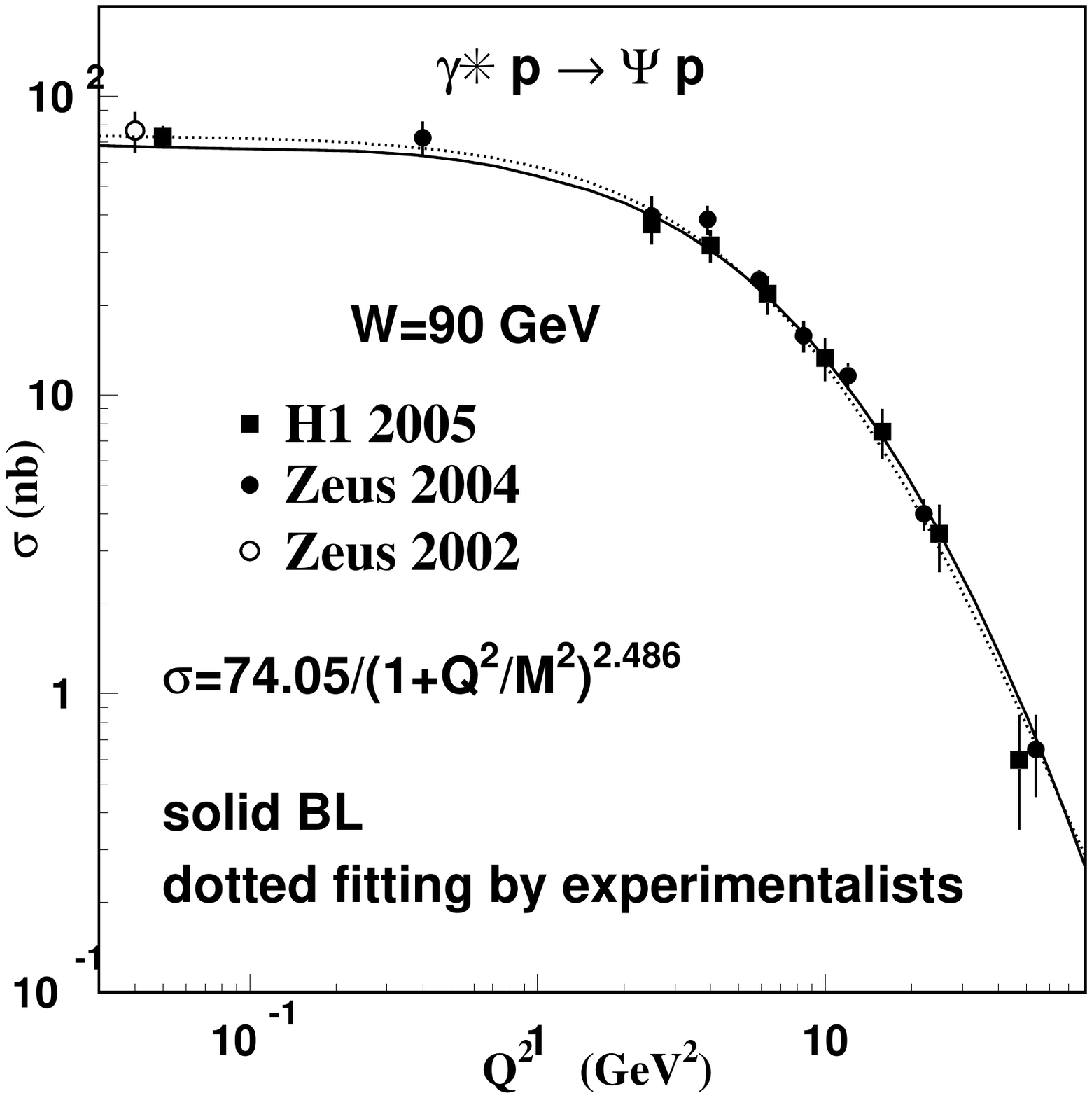}
\includegraphics[height=10.0cm]{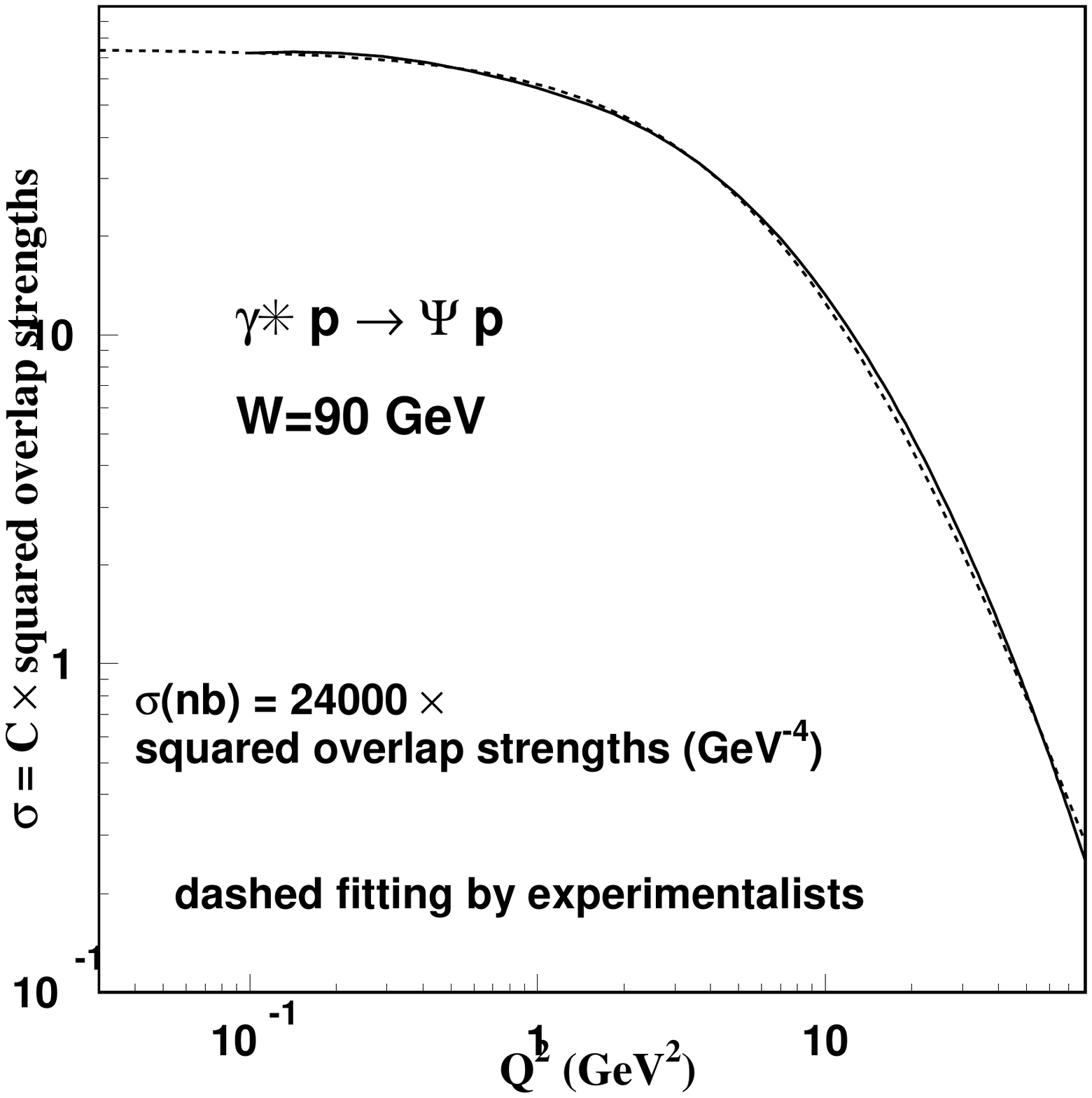}
 \end{figure}
Actually the  $Q^2$ dependence of cross sections is of the form 
$$\sigma=\frac{A_V}{(Q^2+M_V^2)^n}$$
which is  contained in the shape of the integration of products of the 
modified Bessel functions $K_0$ and $K_1$ with powers 
and  Gaussians  that appear in the overlap strengths. 

Good results for cross sections calculated only with these 
quantities are obtained whenever the range of the 
overlap region is small compared to the typical range of 
the nonperturbative interaction governing the process \cite{EFVL}. 

Let us next consider the light vector mesons. Here we may have some 
complications, since it may be necessary to include relativistic and 
higher-order radiative corrections to $f_V$. Moreover, we will need the 
values of $\alpha_S$ for small $m_f$, which are not yet determined 
experimentally.  

While there is a general agreement as to the high energy behavior of the 
strong coupling constant~\cite{Hinchliffe,Prosperi,Bethke}, the infrared 
region, where $\alpha_S(Q^2)$ develops singularities, is not determined. 
Perturbative calculations point to a singular behaviour , but 
the possibility of  finite values at small $Q^2$ has been extensively 
studied in recent years. There are analytical, 
phenomenological and lattice calculations exploring  this possibility, and 
many arguments and techniques have been proposed to solve the problem of the 
elimination of the infrared singularities ~\cite{Prosperi,Natale}.  

A phenomenological proposal to determine the infrared strong coupling constant 
is that of the so called {\em physical coupling} or {\em effective charge}, 
which consists in defining  $\alpha_{S,{\rm eff} }(Q^2)$ from  a given physical 
observable  \cite{Grumberg80}. This was analysed in connection to the 
renormalization-scheme dependence problem in perturbative field theories 
~  \cite{brodskyEFF83,Grumberg84}. This effective charge is a resummation 
in perturbation theory ~\cite{brodskyEFF95,brodskyEFF98,brodskyEFF03}. 
Brodsky et al. argue that a perturbatively calculable physical quantity 
can be used to define an effective charge by incorporating the entire 
radiative correction in its definition ~\cite{brodskyEFF95}.

In line with this phenomenological proposal, for the light vector mesons we 
define an effective strong running coupling constant that absorbs the higher 
order corrections to $f_V$ and so, instead of the corrected
Eq.~(\ref{celmaster}), we write 
\beq \label{effective}
\Ga{(V\rightarrow e^+e^-)} = 
 \Ga_0\left[ 1 - \left(\frac{16}{3\pi}\right) \alpha_{S,{\rm eff} }(m_f)\right]
\enq
Now the reduction factor in Eq.~(\ref{corrected})  is written as
\beq
F(m_f) \approx 
\left[1 - \left(\frac{16}{3\pi}\right) \alpha_{S,{\rm eff}}(m_f) \right]
\enq
and is determined by fitting that leads to universality of cross sections. 
The results obtained are shown in Table \ref{fitted}.
\begin{table}[h]
\label{fitted}
\caption{Reduction factors of $\Gamma$  for $\phi$, $\rho$ and $\omega$ }
\begin{center}
\begin{tabular}{|c c c|} \hline
Meson & $ m_f$(GeV) & $F ({\rm fitted})$\\
\hline
$\phi$ & $0.3$ &  $0.32\pm 0.01$\\
$\rho$ & $0.2$ &  $0.32\pm 0.01$ \\ 
$\omega$ & $0.2$ & $0.29\pm 0.01$ \\ \hline
\end{tabular} 
\end{center}
\end{table}
With this choice for the factors that account for the radiative 
corrections to the electromagnetic decays of the light 
vector mesons, we have the renormalized couplings 
$f_V^{(0)2}$ for all mesons and then we may put the 
electroproduction cross section of all five vector mesons 
in a single line, as shown in Fig. \ref{universal}.  
\begin{figure}
 \caption { Extraction of renormalized $f_V^{(0)2}$ coupling 
factors from the integrated elastic cross sections of electroproduction 
of all vector mesons: theoretical results and 
experimental data converted with same factors for each vector meson, 
as described in the text.}  
\label{universal}
\includegraphics[height=10.0cm]{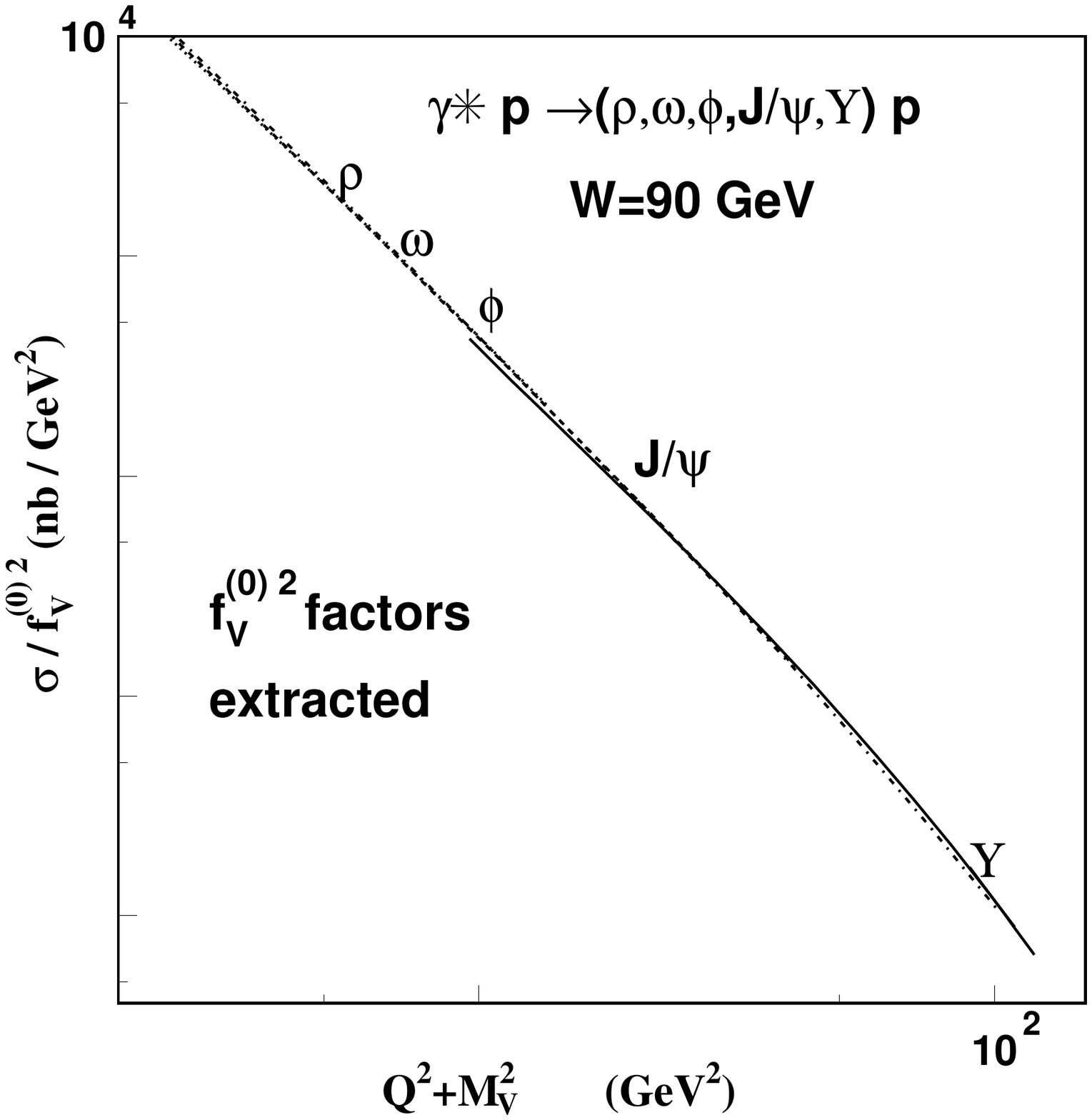}
\includegraphics[height=10.0cm]{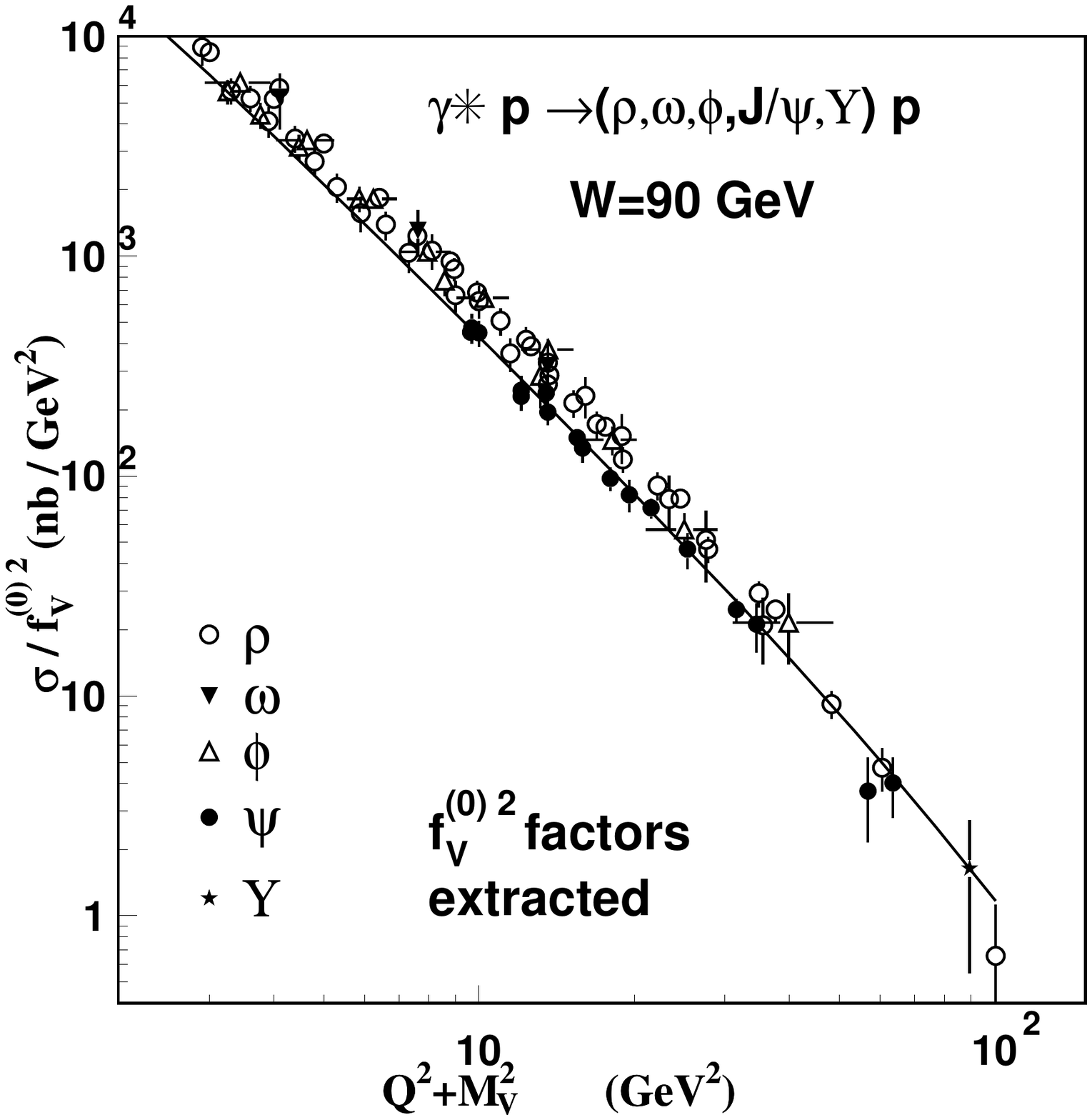}
 \end{figure}
Through the fitted values of $F(m_f)$  we obtain values of the effective 
strong coupling constant at low masses 
\begin{center}
$\alpha_{S,{\rm eff} }(m_f) \approx 0.4$~~ , ~~for $m_f \approx (0.2-0.3)~ \GeV$.
\end{center}
This result is similar to values found in the literature.
  
On the theoretical side, there are many calculations leading to a region below 
$\alpha_S(Q^2) \approx 1$ for low values of~$Q^2$. Let us explicitly mention 
a few of these.

In a classical paper~\cite{Cornwall82}, studying gauge invariant Schwinger-Dyson 
equations, Cornwall advocated the dynamical generation of an 
effective gluon mass, resulting in a gluon propagator that reaches a finite, 
non vanishing value in the deep infrared and in $\alpha_{S}(Q^2)$ which is 
asymptotically free in the ultraviolet and attains a finite infrared value.
This general picture is confirmed by using large volume lattices, 
for SU(2) ~\cite{Cucchieri} and SU(3)~ \cite{Bogolubsky}~pure Yang-Mills. 
Other recent studies are in agreement with Cornwall results
~\cite{Natale,Aguilar06,Aguilar08,Aguilar09}. 

According to the original Cornwall proposal, we have
\beq \label{Cornwall}
\frac{\alpha_S(M_{{Z}_0})}{\alpha_S(Q^2)} =
 1 +  \frac{\alpha_S(M_{{Z}_0})~\beta_0}{2\pi}~\ln \frac{Q^2+4m_g^2}{M_{{Z}_0}} ~. 
\enq
Here, $M_{{Z}_0}$ is the mass of $Z_0$ used as reference, 
$\beta_0 = 11-\frac{2}{3}n_f$, with $n_f = $ number of flavors, and $m_g$ is 
the gluon mass, which can be neglected in face of $M_{{Z}_0}$.

This infrared finite strong coupling constant has been used to study the 
asymptotic pion form factor~\cite{Ji}, resulting in a better agreement 
with the data.

The values of the strong coupling constants that appear 
in the  radiative corrections to the electromagnetic decays, 
which modify the effective couplings,  are exhibited in 
Fig. \ref{alphastrong}. 
\begin{figure}
 \caption { Strong running coupling constant. The dashed lines are the usual 
 perturbative expresions, and the solid line represents Eq. (\ref{Cornwall}) 
with $m_g = 0.6$ GeV .  }  
\label{alphastrong}
\includegraphics[height=10.0cm]{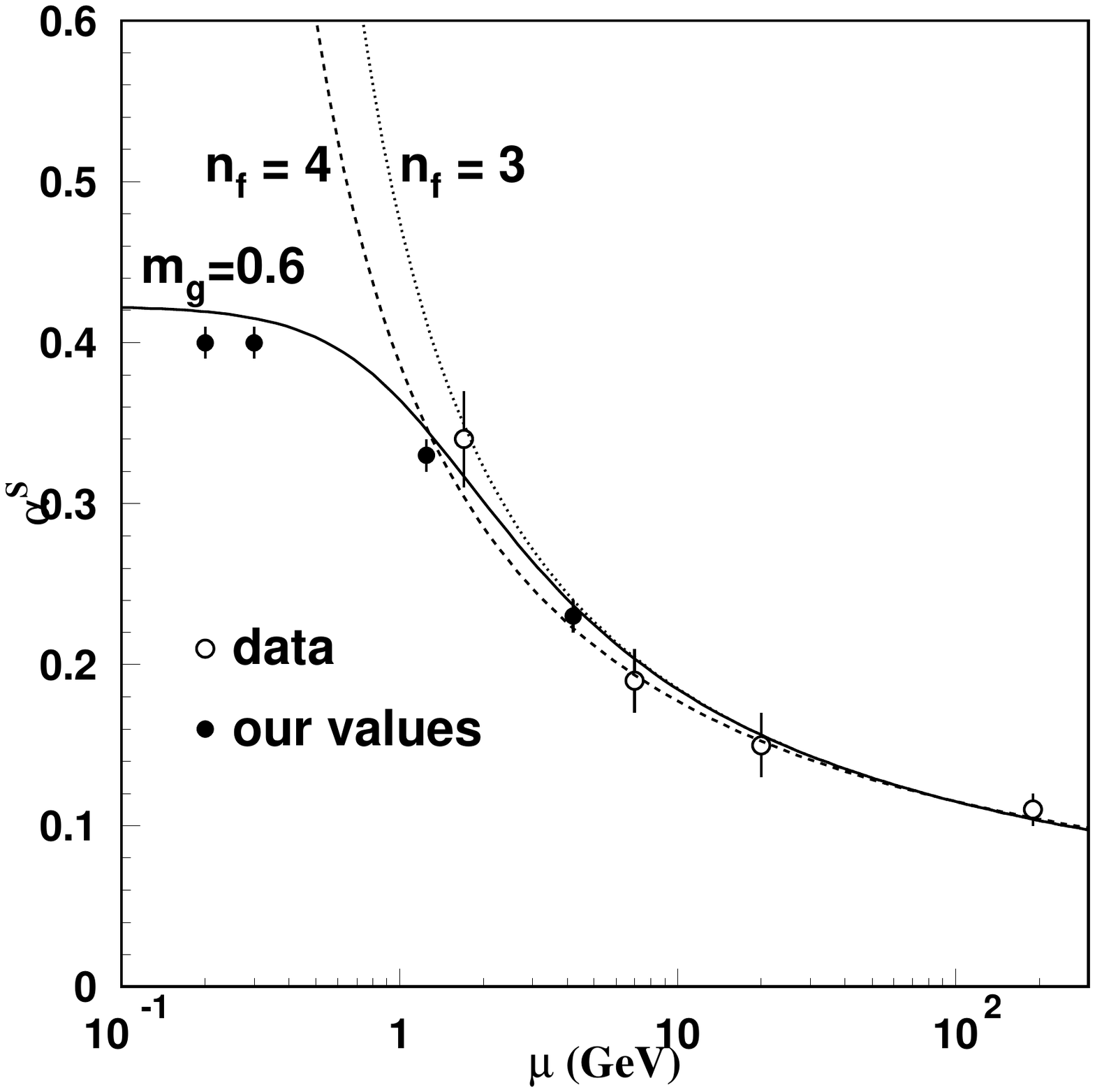}
 \end{figure}
In the figure we compare the values here obtained  with  measurements  
and plot the perturbative and nonperturbative \cite{Cornwall82} 
calculations. In the nonperturbative case we use $m_g$  = 0.6 GeV . 
The figure shows that the value of  $\alpha_{S,{\rm eff } } \approx 0.4$ 
for low masses, obtained by our fitting, is in the range of values 
predicted by Cornwall \cite{Cornwall82} and by more recent work. 
In a later paper \cite{Cornwall07}, and using a different technique 
(the functional Schr\"odinger equation), Cornwall estimated 
$\alpha_S(M^2) \approx  0.4$ for M = 0.6~GeV, which is again in agreement 
with our result. The same is true for  ref.  \cite{Aguilar08}, where 
$\alpha_S$ is found to be a bit below 0.4 for $Q^2 \approx 0.1~\GeV ^2$. 
In a recent paper \cite{Aguilar09}, calculations of QCD effective charges
performed within two different theoretical frameworks are shown to acquire  
a common value $\alpha_S(0)= 0.7 \pm 0.3$.

On the phenomenological side, several different analyses of scattering 
data sensitive to low momentum effects point to  values of  $\alpha_S(Q^2 = 0)$ 
~ in the range~\cite{Natale}. 
\begin{center}
$\alpha_S(0) \approx 0.7 \pm 0.3$
\end {center}
 which is in agreement with our values.

In addition, we mention that the underlying stochastic vacuum model
also predicts finite strong coupling  $\alpha_S \approx 0.8$  in the 
infrared region \cite{SSDP}.

  \section{Conclusions   \label{conclusions} } 

 Our treatment of elastic vector meson electroproduction stresses 
nonperturbative aspects of the QCD calculation, showing 
that large part of the description of this process is supplied by the 
hadronic  structure, namely, the wave functions of photons and vector 
mesons and a simple diquark model for the proton.  
Details of the structure of the proton, that participates in the 
processes like an overall object, determined by its size only, do not 
enter in the description of the integrated observables.

A full description of the framework of the calculations can be
found in previous papers \cite{DFK94,EFVL,DGKP97,DF02,DF03}. 
The calculations, without free parameters, show that quite simple features 
are able to explain the data, covering a wide range of 5 orders of magnitude 
in the cross sections\cite{DF07}.

 Each of the different mesons enter in the calculation characterized only by 
the masses and charges of its quark contents, and with their normalized wave 
function individualized only by the corresponding electromagnetic decay rate 
(related to the value of the wave function at the origin).
The main features of the  $Q^2$ dependence of electroproduction of vector 
mesons are contained in the overlap integral of the light cone wave functions 
of photons and mesons folded with the basic $r^2$ behaviour of the dipole
cross sections   \cite{EFVL} .  
 
In this paper we concentrate on the integrated elastic cross 
sections, studying its universality in the $Q^2$ dependence . 
The energy dependence, which is well described by the two-pomeron model, 
is not detailed here : we choose a value W=90 GeV for which 
experiments are available for all vector mesons.

Our calculation gives a unified description of the electroproduction for 
all five S-wave vector mesons. The $Q^2$ dependence of the integrated 
elastic cross section shows a regular shape when plotted against the variable 
$Q^2+M_V^2$. The lines are almost parallel, and it is well known that,
after extraction of factors equal to the squared electric charges of the 
constituent quarks in the vector mesons (with appropriate averages in 
the $\rho$ and $\omega$ cases), the reduced  cross sections tend to 
concentrate near a common line. We show that this approximate universality 
becomes more exact when the extracted factor is the square of the electromagnetic 
coupling $f_V$ . This improved universality is interesting because 
this factor is more meaningful than the squared charge, since  
it is connected to the value of the wave function at the origin, 
entering in the determination of the basic parameter $\omega$ of the wave 
function, and fixing the size of the vector meson. 

The numerical value of the coupling $f_V$ for each vector meson is fixed  
by the experimental value of the electromagnetic decay rate 
$\Ga_{V\rightarrow e^+e^-}$. This experimental value  is affected by 
radiative corrections at the electromagnetic vertex, which depend 
on the masses of the intervening quarks. We remark that a more 
exact universality would be observed if these corrections were not present.

The criterion of universality of reduced cross sections imposed for  
the light meson cases, using the same expression for the radiative 
correction, leads to effective values for the strong coupling 
constant at the low mass values. These values are in iteresting 
agreement with theoretical studies concerning the infrared 
behavior of the gluon propagator.

The successful description of electroprodution  cross sections,
without free parameters, including all vector mesons, and the 
universality property that is inherent to our nonperturbative 
framework, shows that microscopic details  are not important 
for the observables here studied. Only the dipole distribution 
in the wave functions and the hadronic sizes, and the range of 
the interaction  mediated by the nonperturbative QCD vacuum field, 
are relevant for the phenomenology.

 
\begin{acknowledgements}
 The authors wish to thank 
CNPq (Brazil) and FAPERJ (Brazil) for support of the scientific 
collaboration program between Heidelberg and Rio de Janeiro groups
working on hadronic physics. One of the authors (EF) is grateful to 
CNPq (Brazil) for research fellowship and grant. 
  \end{acknowledgements}

\appendix {Photon and vector meson light-cone wave functions}

\section{Photon and vector meson light-cone wave functions}

	The color part of the photon wave function can be  treated separately and the result is an overall multiplicative factor$\sqrt{N_c}$. The helicity and spatial configuration part of the wave function  $\psi_{\ga^*,\lambda}(Q^2;z,r,\theta)$ is calculated in light-cone perturbation theory. In lowest order, for each 
polarization $\lambda$,  we write 
~\cite{brodsky94,DGKP97,lep80,nik91,bjkgs,lep82}, 
\begin{eqnarray}  \label{ph1}
\psi_{\ga^*,1}(Q^2;z,r,\theta)&=&\hat e_f \frac{\sqrt{6\alpha}}{2 \pi} 
 \Big[ i \ep e^{i\theta}
(z\de_{h,+}\de_{\bar h,-}-\bar z \de_{h,-}\de_{\bar h,+})K_1(\ep r) 
    \nonumber  \\
&& + m_f\de_{h,+}\de_{\bar h,+}K_0(\ep r)\Big] ~ , 
  \end{eqnarray}
  \begin{eqnarray} \label{ph2}
\psi_{\ga^*,-1}(Q^2;z,r,\theta)& =&\hat e_f \frac{\sqrt{6\alpha}}{2 \pi} 
\Big[ i \ep e^{-i\theta}
(\bar z \de_{h,+}\de_{\bar h,-}-z\de_{h,-}\de_{\bar h,+})K_1(\ep r) 
    \nonumber\\
&& + m_f\de_{h,-}\de_{\bar h,-}K_0(\ep r)\Big] 
\end{eqnarray}
and
\beq \label{ph3}
\psi_{\ga^*,0}(Q^2;z,r)=\hat e_f \frac{\sqrt{3\alpha}}{2 \pi} 
 (-2 z \bar z) ~   \de_{h, -\bar h}~ Q~ K_0(\ep r) ~ , 
\enq
where  
\beq  \label{eps}    \ep=\sqrt{z\bar z Q^2+m_f^2} ~ ,\enq
 $\alpha=1/137.036$ and $m_f$ is the quark mass, $\hat {e}_f$ is the quark charge in units of the elementary charge for each flavor $f$; $z (\bar {z})$ is the quark (antiquark) longitudinal momentum fraction; $h (\bar {h})$ is the quark (antiquark) helicity; $K_0$ and $K_1$ are the modified Bessel functions.
In the longitudinal case there is an additional $\delta^2(\vec r)$ which ca be dropped because the color interaction vanishes at zero transverse distance.
The quark masses we use are~$m_u = m_d = 0.2,~ m_s = 0.3,~ m_c = 1.25$~and~$m_b$ = 4.2~(values in GeV).

Meson wave functions are more model dependent than photon wave functions. We take for the vector meson the spin structure from the vector current leading to expressions similar to those of of the photon~\cite{DGKP97,KDP,nik91}.
\begin{eqnarray} \label{VM1}
\psi_{V,+1}(z,r, \theta) &=&  \Big( - i e^{i\theta}
\partial_r(z\de_{h,+}\de_{\bar h,-}-\bar z \de_{h,-}\de_{\bar h,+}) 
                        \nonumber\\
&& + m_f\de_{h,+}\de_{\bar h,+}\Big)\phi_V(z,r) ~ , \nn \\
\psi_{V,-1}(z,r, \theta) &=&  \Big( - i e^{-i\theta}
\partial_r(\bar z\de_{h,+}\de_{\bar h,-}- z \de_{h,-}\de_{\bar h,+})
       \nonumber\\
&& + m_f\de_{h,-}\de_{\bar h,-}\Big)\phi_V(z,r)  
\label{mesontr}\end{eqnarray}
and
\beq   \label{VM3}
\psi_{V,0}(z,r) =
   \Big(\om  4 z \bar z \de_{h, -\bar h}\Big)~\phi_V(z,r) ~ .
\label{mesonlo}\enq

 In this paper we use the functional form for the scalar function ${\phi_V(z,r) }$ proposed by Brodsky and Lepage ~\cite{lep80}

\beq \label{BL}
\phi_{BL}(z,r) = \frac{N}{\sqrt{4 \pi~}}
\exp\Big[-\frac{m_f^2(z-\frac{1}{2})^2}{ 2z\bar{z}\om^2 }\Big]~\exp[-2
z\bar{z}\om^2 r^2] ~ .
\enq
\smallskip
 
	It contains two parameters, $N$ and $\om$, which are determined, as explained in section 2,  by the normalization condition and the leptonic decay width,~$\Ga_{V\rightarrow e^+e^-}$, which is related to $f_V$, the coupling of the vector meson to the electromagnetic current.   Table ~\ref{PDGtab}~ shows vector meson data and the values obtained for $f_V$ through eq.~\ref{fv}.

\begin{table} [h]
 \caption{ \label{PDGtab} S-wave vector meson  data.
 The coupling  $f_V$ and the electromagnetic decay width
 $\Ga_{V\rightarrow e^+e^-}$
 are related through eq.\ref{fv}. 
 The quantity $\hat e_V$ is the effective quark charge in units of
 the elementary charge, determined by the $q\bar q$ structure of
  each meson.  }
 \begin{center}
 \begin{tabular}{|l c c c c |c} \hline
 Meson & $ M_V$(MeV) & $\hat e_V$ &$\Ga_{V\rightarrow e^+e^-}$ (keV) &$f_V$ (GeV) \\
 \hline
 $\rho(770)$  &$775.49\pm 0.34   $ & $1/\sqrt{2} $&$7.04\pm 0.06$&$0.15645\pm 0.0043$\\
 $\omega(782)$&$782.65\pm 0.12 $ & $1/3\sqrt{2}$&$0.60\pm 0.02$&$0.04588\pm 0.0008$\\
 $\phi(1020)$   & $1019.455\pm 0.020$ & $-1/3     $&$1.27\pm 0.04$&$0.07619\pm 0.0012$\\
 $J/\psi(1S)$ & $3096.916\pm 0.011$ & $2/3       $&$5.55\pm 0.141$&$0.27759\pm 0.0035$\\
 $\Upsilon(1S)$& $9460.30\pm 0.26$ & $-1/3      $&$1.340\pm 0.018$&$0.23839\pm 0.0016$\\
 \hline
 \end{tabular}
 \end{center}
 \end{table}
Numerically,we use $f_V^T = f_V^L = f_V$, with $f_V$ calculated through eq.~\ref{fv}. The values thus obtained, together with other vector meson data, are presented in Table~\ref{PDGtab}. Table~\ref{WFparam} shows our results for~$\om$,~$N$,~and radius $r$ for each meson in each state, calculated with eqs.~\ref{fV1}~and~\ref{fV0}~, the wave function normalization equation $\int_0^1 dz~\int d^2 {\mathbf r}~|\psi_{V,\lambda}(z,\mathbf r)|^2 = 1$~ and the mean value of $r$.
\begin{table} [h]
  \caption{ \label{WFparam}  Parameters of the vector meson wave functions }
 \begin{center}
 \begin{tabular}{|l c c c c c c|}\hline
\multicolumn{1}{|l}{} &\multicolumn{3}{c}{\bf Transverse} &
          \multicolumn{3}{c}{\bf Longitudinal}\\
&$~ \om$(GeV)& $N$ &radius (fm)   &$~ \om$(GeV)& $N$ &radius(fm)  \\  \hline
$\rho(770)$	&$0.2809 $  &$2.0820$&$1.0637$&$0.3500$   &$1.8366$&$0.6402$ \\
$\omega(782)$	&$0.2618 $  &$2.0470$&$1.1110$&$0.3088$   &$1.8605$&$0.7194$\\
$\phi(1020)$	&$0.3119 $  &$1.9201$&$0.8569$&$0.3654$   &$1.9191$&$0.5973$ \\
$J/\psi(1S)$	&$0.6452 $  &$1.4752$&$0.3418$&$0.7140$   &$2.2769$&$0.2898$ \\	
$\Upsilon(1S)$	&$1.3333 $  &$1.1816$&$0.1551$&$1.3851$   &$2.7694$&$0.1456$ \\
\hline  \end{tabular}  \end{center}  \end{table}

After summation over the helicity indices, the overlaps of the photon and vector meson wave functions (eq.~\ref{overlap}) are

\begin{eqnarray} \label{overBL1}
 \rho_{\ga^* V ,\pm1}(Q^2;z,r)=\hat e_V\frac{\sqrt{6\alpha}}{2\pi}
 \Big(4 \ep_f ~\om^2 rz\bar{z} \big[z^2+\bar{z}^2\big] K_1(\ep_f ~r)
    + m_f^2 K_0(\ep_f ~r)\Big)~\phi_{BL}(z,r), 
 \end{eqnarray}
for the transverse case, 

\beqa \label{overlong}
 \rho_{\ga^* V ,0}(Q^2;z,r) = -16 \hat e_V\frac{\sqrt{3\alpha}}{2\pi}~ \om~
     z^2 \bar{z}^2~Q~ K_0(\ep_f ~r) \phi_{BL}(z,r),
 \enqa
for the longitudinal case.

The overlap strengths, formed by integration
 over the internal variables of the quark-antiquark pairs
 of the overlap function multiplied by $r^2$ are 
 \beq  \label{str1meson}
 Y_{\ga^* V,T} (Q^2)=
  \int_0^1 dz \int d^2{\mathbf r} ~ r^2 ~
 \rho_{\ga^*V,\pm 1}(Q^2;z,r)
     \equiv \hat e_V ~ \widehat{Y}_{\ga^* V,T} (Q^2)
 \enq
for transverse polarization and
\beq \label{str0meson}
 Y_{\ga^* V ,L} (Q^2)=
  \int_0^1 dz \int d^2{\mathbf r} ~ r^2 ~
 \rho_{\ga^* V,0}(Q^2;z,r)
  \equiv \hat e_V ~ \widehat{Y}_{\ga^* V,L} (Q^2)
 \enq
for longitudinal polarization.

\end{document}